\documentclass{cs19proc}

\editors{G.~A. Feiden}
\publisher{Zenodo}
\conference{The 19th Cambridge Workshop on Cool Stars, Stellar Systems, and the Sun}
\conferencedate{2016}

\title{(Re)solving Mysteries of Convection and Mass Loss of AGB Stars: 
What New Models and Observations Tell Us About Long-Standing Problems}
\author{Susanne H{\"o}fner} 

\affiliation{
Department of Physics and Astronomy, Division of Astronomy and Space Physics, Uppsala University, Box 516, SE-75120 Uppsala, Sweden
}

\abs{The recent progress in high-spatial-resolution techniques, spanning wavelengths from the visual to the radio regime, is leading to new valuable insights into the complex dynamical atmospheres of Asymptotic Giant Branch (AGB) stars and their wind forming regions. Striking examples are images of asymmetries and inhomogeneities in the photospheric and dust-forming layers which vary on time-scales of months. These features are probably related to large-scale convective flows predicted by 3D `star-in-a-box' models. Furthermore, high-resolution observations make it possible to measure dust condensation distances, and they give information about the chemical composition and sizes of dust grains in the close vicinity of cool giants. These are essential constraints for building realistic models of wind acceleration and developing a predictive theory of mass loss for AGB stars, which is a crucial ingredient of stellar and galactic chemical evolution models.}

\begin{document}

\maketitle

\section{Introduction}

\begin{figure}[ht]
	\centering
	\includegraphics[width=8.5cm]{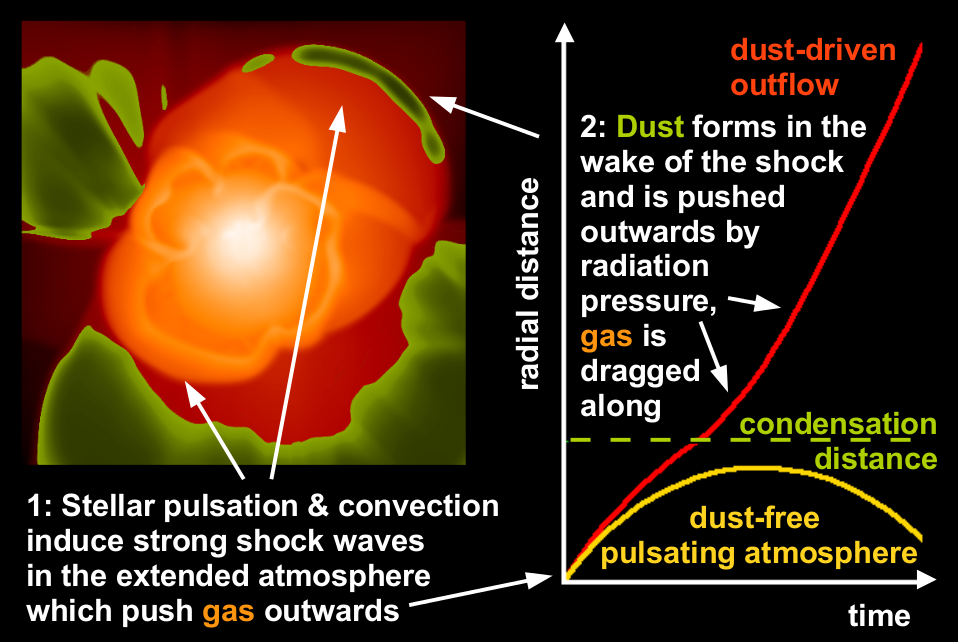}
	\caption{An illustration of the two-step wind mechanism. 
{\em Left:} A slice through a 3D 'star-in-a-box' model \citep{frey08}, showing effects of convection and pulsation on the gas density (orange) and dust distribution (green) in the extended atmosphere of the star (width of the box: about 4 $R_{\ast}$). 
{\em Right:} Schematic diagram showing trajectories of mass layers in the atmosphere and wind acceleration zone.}
	\label{f_PEDDRO}
\end{figure}

\noindent
Asymptotic Giant Branch (AGB) stars represent a short but decisive phase in the late evolution of stars that start out with about 0.8--8 solar masses \citep[see, e.g.,][or the presentation by Karakas, this conference]{LaKa16}. Stellar winds with typical velocities of 5--20 km/s are a defining feature of these cool giant stars, affecting both their observable properties and their final fate. The heavy mass loss due to these outflows (typically $10^{-8}$--$10^{-4}$ $M_{\odot}$/yr, i.e. about 6--10 orders of magnitude higher than for the sun) enriches the interstellar medium with newly-produced elements and dust, and eventually turns low- and intermediate mass stars into white dwarfs. 
The mass loss is commonly assumed to be caused by a two-step process: atmospheric levitation due to strong shock waves, followed by radiative acceleration of dust grains which form in the cool extended atmosphere and drag the surrounding gas along (see Fig.~\ref{f_PEDDRO}). The shock waves are presumably triggered by a combination of large-scale convective flows and stellar pulsations, which are associated with a strong variability of the stellar luminosity with typical periods of 100 -- 1000 days (so-called Mira stars and semi-regular long-period variables). 

Until very recently, observational studies of dynamical processes in cool giants had to rely mostly on indirect methods, using photometry and high-resolution spectroscopy. Observed light curves and changes in spectral line profiles due to Doppler shifts hold valuable information about stellar pulsations, convective flows, and wind properties, but they need to be decoded using assumptions about the physical structure of the observed stars. The current progress in high-spatial-resolution techniques, spanning wavelengths from the visual to the radio regime, adds a new dimension to this picture. Instead of having to speculate about the structure of unresolved light sources, the evolving dynamical patterns on the surfaces and in the dusty atmospheres of the most nearby cool giants can now be observed directly and in real time \citep[e.g.,][]{ohnaka15,Haubois15,ohnaka16,khouri16}. 
Combining such observations with self-consistent models of dynamical processes in the stellar interior (convection, pulsation), the atmosphere (propagating shock waves) and the inner wind region (dust formation, radiative acceleration) is a promising way to understand the fundamental properties of cool giants and their impact on stellar and galactic evolution. 

\begin{figure*}[ht]
	\centering
	\includegraphics[width=18cm]{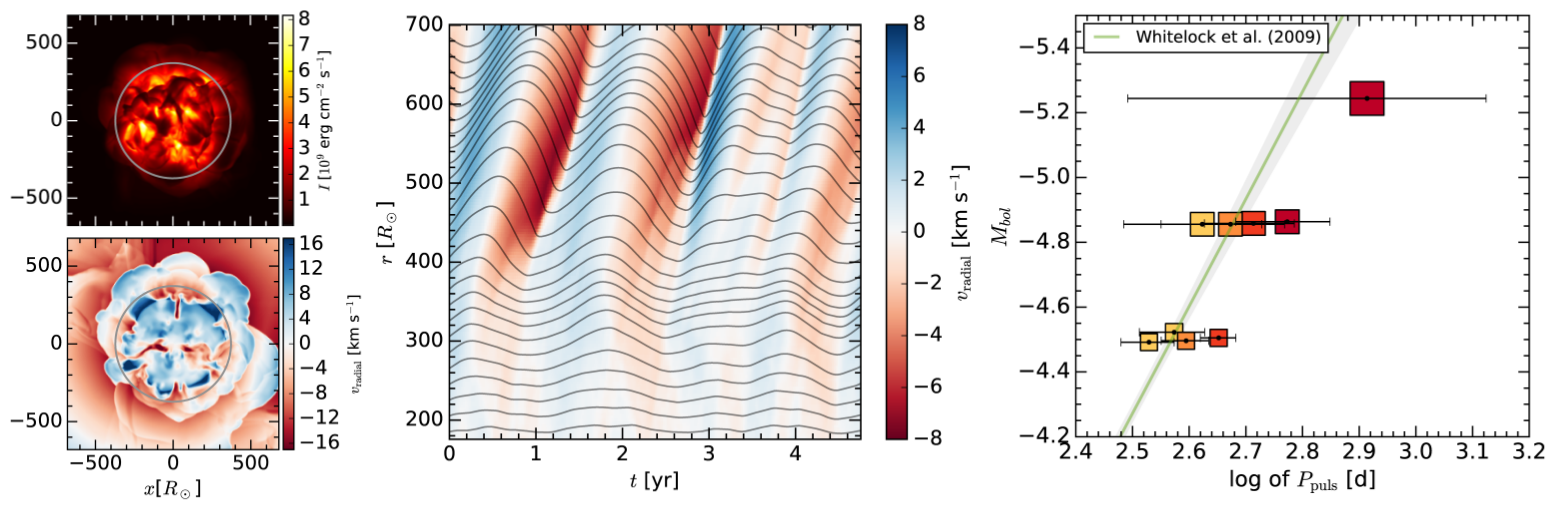}
	\caption{Properties of 3D star-in-a-box models by Freytag, Liljegren \& H{\"o}fner (2016, submitted to A\&A): 
{\em Top left panel:} A snapshot of the surface intensity. 
{\em Lower left panel:} A snapshot of he radial component of the gas velocity for a slice through the center of the star, illustrating the size scales of convective flows. 
{\em Middle panel:} Radial gas velocity averaged over spherical shells, plotted as a function of depth and time. Radial pulsations, not obvious from looking at the radial velocity slices alone, are clearly visible as alternating expansions (blue) and contractions (red). The stellar interior (below about 400 $R_{\odot}$) shows a velocity pattern similar to standing waves. They trigger propagating shock waves in the atmosphere, which move upwards as time proceeds. The black lines trace the expansion and contraction of the stellar interior during a pulsation cycle, and the ballistic movements of gas in the atmosphere, triggered by the shock waves. 
{\em Right panel:} Mean luminosities of the models plotted against their (range of) pulsation period(s), extracted from the average radial velocities (as shown in the middle panel) by Fourier analysis. The current models feature the following parameters: luminosities of 5000, 7000 or 10000 $L_{\odot}$, combined with effective temperatures of about 2500--2800~K (dark red to yellow symbols), and a mass of 1 $M_{\odot}$. They show good agreement with the observed period-luminosity relation by \citet[][]{whitelock09}, indicating that they successfully capture the main features of interior dynamics.}
	\label{f_3D_flh16}
\end{figure*}

The presentation of recent results given below follows the flow of matter from the stellar interior to the dusty circumstellar envelope, confronting models and observations for the different regions. 
Due to the fast developments taking place at present, the intention is to highlight a few key examples, rather than to give an exhaustive overview. 
More examples of recently published and on-going work in this field can be found in the summary of the splinter meeting on Mass Losing Asymptotic Giant Branch Stars and Supergiants (Whitelock {\it et al.}, this conference).

\section{Convection and atmospheric structures}\label{s_3Dmod}

\noindent
3D radiation-hydrodynamical (RHD) models of solar convection have a long and successful history of  reproducing the observed granular patterns and improving abundance determinations 
\citep[e.g.,][]{dravins81,nordlund82,ludwig09,magic13}.
Corresponding models of AGB stars, however, are still at an exploratory stage, both due to the large computational effort involved and the intricate physics of these stars \citep[see, e.g.,][]{freytag12}. When stars evolve away from the main sequence, turning into cool luminous giants, convection develops from small-scale surface `granules', as seen on our sun, to giant convection cells which reach deeply into the stellar interior. Instead of studying a small part of a surface convection zone with local models, as is appropriate for solar-type stars, the large convection cells of cool giants require global `star-in-a-box' models. 

According to the 3D RHD models of \citet{frey08}, the large-scale convective flows in AGB stars produce distinctive, variable patterns in surface intensity, and they trigger strong radiative shock waves which propagate outwards through the extended atmospheres. They induce complex density structures which leave their imprints on circumstellar dust distributions, since grain growth is sensitive to both prevailing densities and temperatures (see Fig.~\ref{f_PEDDRO}). 

Observing visible light scattered by dust grains in the close vicinity of the AGB star W Hya with VLT/SPHERE-ZIMPOL, \citet[][]{ohnaka16} recently obtained resolved images of three clumpy dust clouds at about 2--3 stellar radii, which are comparable to structures resulting from large-scale convection in the models of \citet{frey08}. The analysis of the observed polarized intensity maps indicates that they trace an optically thin medium with density enhancements of about a factor of 4 in the clouds, and grain radii of 0.4--0.5 microns. The latter supports a scenario of wind-driving by photon scattering on near-transparent dust grains \citep[][]{hoefner08}, which will be further discussed in Sect.~\ref{s_wind}. 

Using the same instrument, \citet[][]{khouri16} obtained high-angular-resolution images of the AGB star R Dor at two epochs, demonstrating that the size and morphology of surface patterns change significantly in less than 2 months, which is compatible with convection-induced structures found in the 3D models. They interpret these changes as being due to variability in the excitation and/or density of the TiO molecule. They also observe light scattered by dust grains close to the star, and a strong decrease of the dust density further out which may point to wind acceleration, or a variable mass loss rate. 

In addition to these extreme adaptive optics imaging studies with VLT/SPHERE, the surfaces and inner atmospheres of several nearby AGB stars have recently been mapped with VLTI/PIONIER in the near-IR (H-band), e.g. the Mira star R Car \citep[][]{monnier14}, the carbon-rich Mira R For (Paladini {\it et al.,} in prep.) and the carbon-rich AGB star R Scl (Wittkowski {\it et al.,} in prep.), famous for the large-scale spiral structure in the circumstellar envelope discovered with ALMA \citep[][]{maercker12}. The reconstructed images of these stars show intensity patterns in qualitative agreement with convection models, and strong effects of dust absorption for the carbon-rich objects. Other instruments that have been used to map asymmetries in the molecular layers are VLTI/AMBER \citep[e.g.,][]{Haubois15}, the Visible and Infrared Mapping Spectrometer (VIMS-IR) aboard the Cassini spacecraft \citep[by watching the star pass behind Saturn's rings, e.g.,][]{stew16}, or ALMA. Based on recent ALMA data, \citet[][]{kaminski16} found a clumpy, broken-ring structure for the distribution of AlO at about 2.5 stellar radii around Mira. This molecule may be a precursor of Al$_2$O$_3$ grains, which are candidates for explaining the scattered visible light in the close vicinity of W Hya. 

\section{Convection and pulsations of AGB stars}

\noindent
Long before the surfaces, or even the extended atmospheres, of AGB stars where resolved, the dynamical nature of these cool giants was inferred from their visual and near-IR light curves. The pronounced changes in radiative flux are generally attributed to large-amplitude pulsations, but the underlying physical mechanisms are still not well understood. A strong motivation to solve this long-standing problem is the supposed key role of pulsation for the wind mechanism, implied by observed correlations of mass loss indicators and pulsation properties \citep[e.g.,][]{McDon16}. 

Over the last 10--15 years, significant progress has been made in identifying several of the period-luminosity sequences found in survey data (e.g., MACHO, OGLE) with different pulsation modes, using linear, non-adiabatic models \citep[e.g.,][]{wood15}. Mira stars are interpreted as fundamental-mode (radial) pulsators, whereas semi-regular AGB stars are probably pulsating in the first overtone. The improvements in data have also led to the discovery of substructures in the P-L sequences, probably due to non-radial modes. There are, however, open problems, e.g., regarding the understanding of long secondary periods \citep[e.g.,][]{saio15}, or the relations between first overtone and fundamental mode periods when interpreting them as an evolution sequence (Trabucchi {\it et al.}, this conference). The later problem cannot be resolved by using non-linear radial fundamental-mode pulsation models \citep[][]{ireland08,ireland11} instead of linear ones, but Trabucchi and collaborators find a smaller discrepancy between models and observations in that case.  

\begin{figure}[ht]
	\centering
	\includegraphics[width=8.5cm]{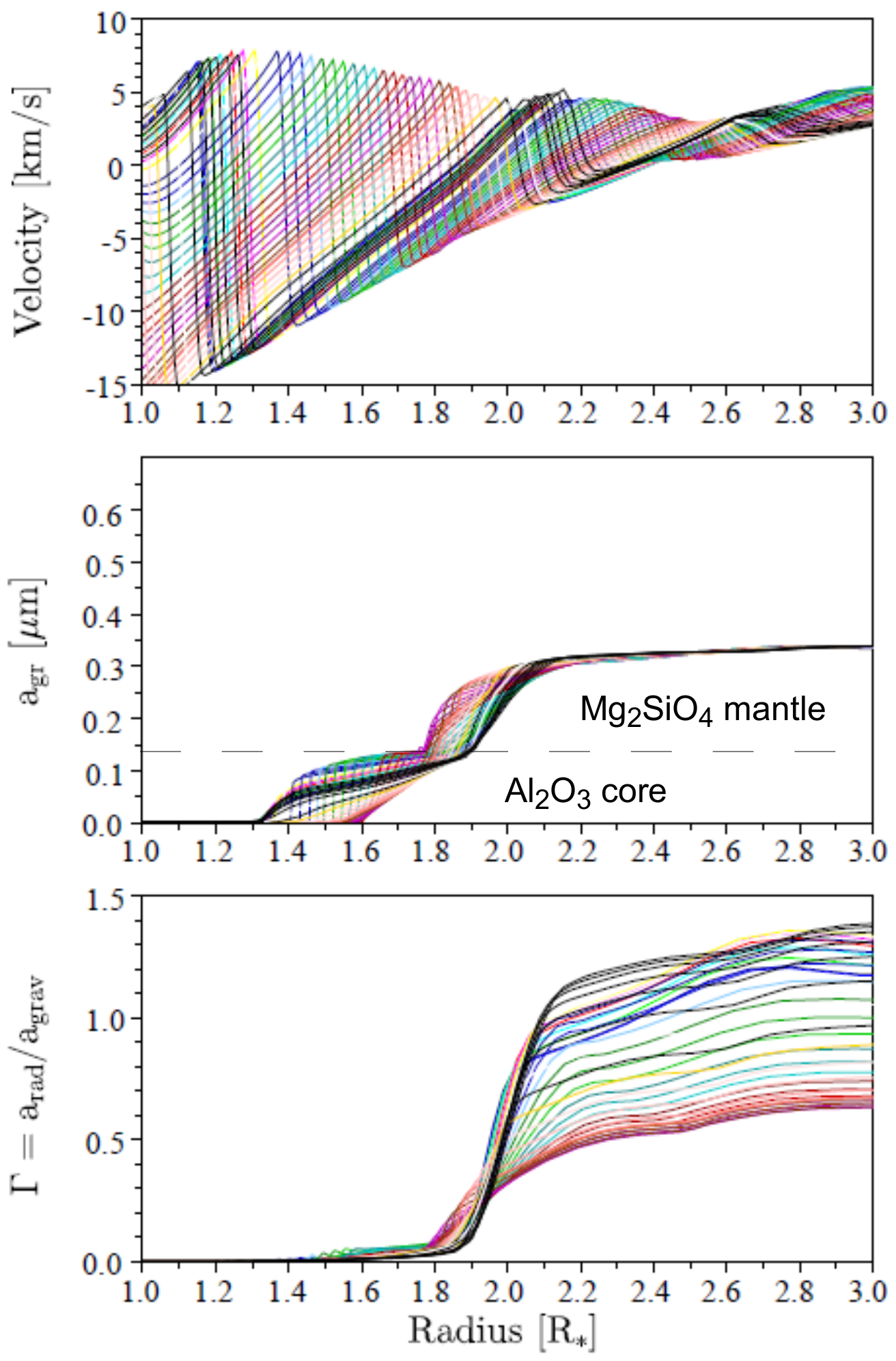}
	\caption{Time-dependent radial structure of a dust-driven wind model by \citet[][]{HBAA16arx}, zoomed in on the dust formation region (snapshots of 40 pulsation phases). {\em Top to bottom}: Velocity, radius of the composite grains, and ratio of radiative to gravitational acceleration. See text for details.}
	\label{f_cmg_str}
\end{figure}

An alternative approach to pulsations of AGB stars are global 3D RHD models, as discussed in Sect.~\ref{s_3Dmod}. In this case it is possible to avoid a parameterization of convection (e.g., in terms of mixing length theory) as done in classical linear and non-linear pulsation models, because the large-scale convective flows are an intrinsic part of the models. A first exploratory grid of such simulations for AGB stars was recently presented by Freytag, Liljegren \& H{\"o}fner (2016, submitted to A\&A), demonstrating how convection patterns and pulsation properties depend on stellar parameters (see Fig.~\ref{f_3D_flh16}). Compared to the simulations of \citet{frey08}, the accuracy and stability of the CO5BOLD code has been improved \citep[see][]{freytag13}. The new models also feature a higher spatial resolution, refining the picture of dynamic atmospheric structures which result from a combination of long-lasting giant convection cells surrounded by turbulent downdrafts, short-lived smaller surface granules, and radial fundamental-mode pulsations. 
The radial pulsations show a correlation of periods and luminosities in good agreement with observations, as illustrated in Fig.~\ref{f_3D_flh16}. 

Sound waves, produced in the deeper layers by convection and pulsation, steepen into shocks when reaching the atmosphere. There, they interact and merge, leading to a transition from a fine shock network close to the surface of the convection zone to almost global, more or less radially expanding shock fronts in the outer atmospheric layers. 
These large-scale, almost radially propagating shock fronts create local conditions which are probably similar to those caused by spherical shock waves resulting from pure radial pulsation \citep[for a discussion see, e.g.,][]{witt16}. This is a likely explanation why earlier IR interferometric measurements of shock-induced density structures \citep[e.g.,][]{tej03,weiner04} could be successfully interpreted with spherical models, and why spherical dust-driven wind models give mass loss rates, spectra and photometric variations that agree well with observations 
\citep[e.g.,][]{wljhs00,wach02,jeong03,hoefner_etal03,nowo10,matt10,erik14,bladh15}.

\section{Dust formation and wind acceleration}\label{s_wind}

\noindent
The mechanism which drives the massive winds of AGB stars presumably involves a combination of atmospheric levitation by shock waves and radiation pressure on dust, as illustrated in Fig.~\ref{f_PEDDRO}. Stellar pulsations and large-scale convective flows trigger strong radiative shocks which intermittently push parts of the upper atmospheric layers out to a few stellar radii, creating temporary reservoirs of cool dense gas where solid particles may condense. The dust grains are accelerated away from the star by interactions with stellar photons and drag the surrounding gas along due to frequent collisions. 

This two-step wind mechanism is supported by various types of observations, and not the least by high-resolution spectroscopy which allows to trace gas velocities from the photosphere to the wind region via Doppler shifts in line profiles \citep[e.g.][]{HHR82,SW00,nowo10}. 
The observed outflows show clear signs of dust formation at about 2--5 stellar radii, but the chemical composition, sizes, and optical properties of the grains which are actually responsible for driving the winds have been a long-standing matter of debate \citep[e.g.,][]{hoefner09}. 
Magnesium-iron silicates seem to be good candidates, based on elemental abundances and the prominence of silicate features in mid-IR spectra of AGB stars. Detailed models, however, show that silicates have to be basically Fe-free in the layers where the wind originates, in order to avoid destruction by radiative heating \citep[][]{woit06b}. The resulting low levels of absorption at visual and near-IR wavelengths lead to a low radiative pressure, insufficient to drive an outflow. 

\begin{figure}[ht]
	\centering
	\includegraphics[width=8.5cm]{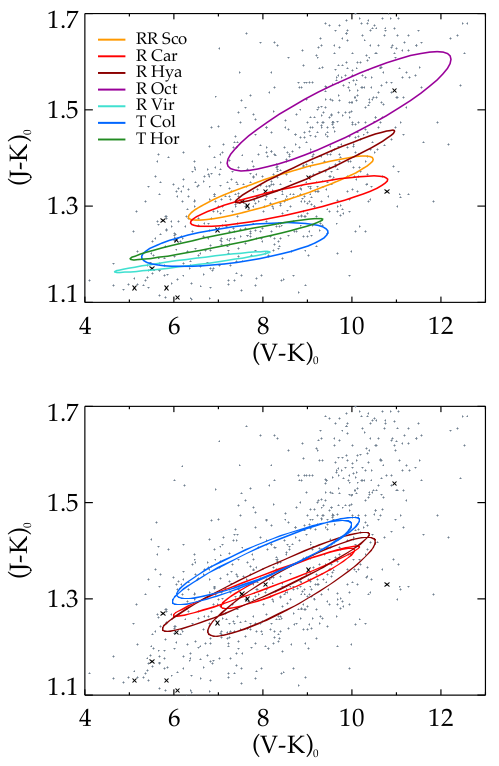}
	\caption{Observed and synthetic photometric variations of M-type AGB stars. {\em Upper panel:} A sample of well-observed stars with regular light curves. {\em Lower panel:} DARWIN models featuring winds driven by composite dust grains with an Al$_2$O$_3$ core and a Mg$_2$SiO$_4$ mantle, as shown in Fig.~\ref{f_cmg_str} \citep[for details see Sect.~\ref{s_wind} and][]{HBAA16arx}.}
	\label{f_cmg_col}
\end{figure}

Therefore, \citet[][]{hoefner08} proposed an alternative mechanism, i.e. wind-driving by scattering of stellar photons on Fe-free silicate grains. This scenario requires grain sizes of about 0.1--1 micron, in order to make the dust particles efficient at scattering radiation in the near-IR wavelength region where the stellar flux peaks. Recent detections of dust grains with sizes of 0.3--0.5 microns in the close vicinity of several cool giants \citep[e.g.][]{norris12,ohnaka16} lend strong support to this scenario. Furthermore, detailed atmosphere and wind models based on this mechanism show good agreement with observations regarding mass loss rates and wind velocities, as well as visual and near-IR colors, and their variation with pulsation phase \citep[][]{bladh13,bladh15}. 

In a recent paper \citet[][]{HBAA16arx} take these studies a step further, including Al$_2$O$_3$ as an additional dust species in the DARWIN models. Spectro-interferometric observations indicate that this material forms at distances of about 2 stellar radii or less, prior to silicate condensation \citep[e.g.][]{witt07,zhao12,karo13}. Al$_2$O$_3$ has been discussed as a possible alternative to silicates as a source of the scattered light observed close to several AGB stars \citep[e.g.][]{ireland05,norris12,ohnaka16}, and as potential seed particles for the condensation of silicates further out in the atmosphere where lower temperatures prevail \citep[e.g.][]{KoSo97a,KoSo97b}. 

\citet[][]{HBAA16arx} conclude that condensation of Al$_2$O$_3$ at the close distances and in the high concentrations implied by observations requires high transparency of the grains in the visual and near-IR region to avoid destruction by radiative heating. This puts an upper limit on the imaginary part of the refractive index, $k$, in the visual and near-IR that is 1--2 orders of magnitude below the values of \citet[][]{koike95}, but similar to a sample of spinel studied by \citet[][]{zeidler11}. Radiation pressure due to Al$_2$O$_3$ grains is found to be too low to drive a wind for typical stellar parameters. This fits well with the scenario of Al$_2$O$_3$ forming a dense, gravitationally bound dust shell at less than 2 stellar radii, as discussed by \citet[][]{ireland05} and \citet[][]{khouri15,khouri16}. However, the formation of composite grains with an Al$_2$O$_3$ core and a silicate mantle can speed up grain growth (see Fig.~\ref{f_cmg_str}), increasing both mass loss rates and wind velocities compared to outflows driven by pure Fe-free silicate grains. Furthermore, these core-mantle grain models lead to variations of visual and near-IR colors during a pulsation cycle which are in even better agreement with observations (see Fig.~\ref{f_cmg_col}). 

Finally, it should be mentioned that a recent study of carbon-rich AGB stars in the Small Magellanic Cloud by \citet[][]{nanni16} favors amorphous carbon grains with sizes well below 0.2 microns, based on simultaneous fitting of several IR colors. In contrast, \citet[][]{maercker10} and \citet[][]{olofsson10} found that carbon grains with radii of 0.15--0.2 microns reproduce scattered light observations of detached shells around galactic carbon stars. This discrepancy might be due to several factors, e.g., the different metallicities of the stars, or that the detached shells where produced during a He-shell flash.  
In any case, it should be noted that even small carbon grains can drive winds, since they efficiently absorb stellar light (in contrast to near-transparent Fe-free silicate grains in stars with C/O $<$ 1, as discussed above). Photon scattering on large carbon grains may increase radiative pressure and boost wind acceleration \citep[][]{matt11}, but it is probably not a vital ingredient of the mass loss mechanism in C-rich AGB stars. 

\section*{Acknowledgments}
\noindent
{I thank Bernd Freytag and Sofie Liljegren for supplying figures of the 3D models which are partly still unpublished. Our work is supported by the Swedish Research Council.}

\bibliographystyle{cs19proc}
\bibliography{hoefner}

\end{document}